\documentclass[nobibnotes,amsmath,aps,amssymb,prl,twocolumn,showpacs,superscriptaddress,groupedaddress]{revtex4}  

\usepackage{caption}
\usepackage{ragged2e}
\usepackage{graphicx}  
\usepackage{dcolumn}   
\usepackage{amssymb}   
\usepackage[utf8]{inputenc}
\usepackage{lscape}
\usepackage{booktabs}
\usepackage{mathptmx, courier, pifont}
\usepackage[scaled=0.92]{helvet}
\usepackage[T1]{fontenc}
\usepackage{textcomp}
\usepackage{rotating}
\usepackage{verbatim}
\usepackage{picinpar}
\usepackage{wrapfig}
\usepackage{multirow}
\usepackage[normalem]{ulem}
\usepackage[colorlinks=true, citecolor=blue, urlcolor=blue, linkcolor=blue,breaklinks=true]{hyperref}
\usepackage{orcidlink}
\usepackage{flushend}
\PassOptionsToPackage{unicode}{hyperref}
\PassOptionsToPackage{naturalnames}{hyperref}
\usepackage{parskip} 
\begin{document}

\title {$\beta$-delayed spectroscopy of $^{80}$Ge$_{48}$ and competition between Gamow-Teller and first-forbidden transitions in $^{80g+m}$Ga$_{49}$ $\beta$ decay}

\author{{R. Li}\orcidlink{0000-0002-2782-2333}}
\email[Contact author: ]{liren824@gmail.com}
\affiliation{Université Paris-Saclay, CNRS/IN2P3, IJCLab, 91405 Orsay, France}
\affiliation{KU Leuven, Instituut voor Kern- en Stralingsfysica, Celestijnenlaan 200D, B-3001 Leuven, Belgium}
\author{{D. Verney}\orcidlink{0000-0001-7924-2851}}
\affiliation{Université Paris-Saclay, CNRS/IN2P3, IJCLab, 91405 Orsay, France}
\author{{I. Matea}\orcidlink{0000-0002-8123-150X}}
\affiliation{Université Paris-Saclay, CNRS/IN2P3, IJCLab, 91405 Orsay, France}
\author{{M. N. Harakeh}\orcidlink{0000-0002-7271-1712}}
\affiliation{ESRIG, University of Groningen, Zernikelaan 25, 9747 AA Groningen, The Netherlands}
\author{{C. Delafosse}\orcidlink{0000-0001-5717-2426}}
\affiliation{Université Paris-Saclay, CNRS/IN2P3, IJCLab, 91405 Orsay, France}
\affiliation{Department of Physics, Accelerator Laboratory, P.O. Box 35, University of Jyväskylä, FI-40014 Finland}
\author{{F. Didierjean}\orcidlink{0009-0002-4950-3162}}
\affiliation{Institut Pluridisciplinaire Hubert Curien, CNRS/IN2P3 and Université de Strasbourg, Strasbourg, France}
\author{{S. Ebata}\orcidlink{0000-0003-3523-7772}}
\affiliation{Graduate School of Science and Engineering, Saitama University, Saitama, Japan}

\author{{L. A. Ayoubi}\orcidlink{0000-0002-1441-9094}}
\affiliation{Université Paris-Saclay, CNRS/IN2P3, IJCLab, 91405 Orsay, France}
\affiliation{Department of Physics, Accelerator Laboratory, P.O. Box 35, University of Jyväskylä, FI-40014 Finland}
\author{{H. Al Falou}\orcidlink{0009-0002-0516-498X}}
\affiliation{Faculty of Sciences 3, Lebanese University, Michel Slayman Tripoli Campus, Ras Maska 1352, Lebanon}
\author{{G. Benzoni}\orcidlink{0000-0002-7938-0338}}
\affiliation{INFN, Sezione di Milano, Dipartamiento di Fisica, Milano, Italy}
\author{{F. Le Blanc}}
\affiliation{Université Paris-Saclay, CNRS/IN2P3, IJCLab, 91405 Orsay, France}
\author{{V. Bozkurt}\orcidlink{0000-0003-4651-0447}}
\affiliation{Department of Physics, Science Faculty, Nigde Omer Halisdemir University, 51240 Nigde, Turkey}
\author{M. Ciemała}
\affiliation{Institute of Nuclear Physics, Polish Academy of Sciences, Krakow, Poland}
\author{{I. Deloncle}}
\affiliation{Université Paris-Saclay, CNRS/IN2P3, IJCLab, 91405 Orsay, France}
\author{M. Fallot}
\affiliation{Subatech, CNRS/IN2P3, Nantes, EMN, F-44307, Nantes, France}
\author{C. Gaulard}
\affiliation{Université Paris-Saclay, CNRS/IN2P3, IJCLab, 91405 Orsay, France}
\author{{A. Gottardo}\orcidlink{0000-0002-0390-5767}}
\affiliation{Laboratori Nazionali di Legnaro, I-35020 Legnaro, Italy}
\author{{V. Guadilla}\orcidlink{0000-0002-9736-2491}}
\altaffiliation[Present address: ]{Faculty of Physics, University of Warsaw, 02-093 Warsaw, Poland.}
\affiliation{Subatech, CNRS/IN2P3, Nantes, EMN, F-44307, Nantes, France}
\author{{J. Guillot}}
\affiliation{Université Paris-Saclay, CNRS/IN2P3, IJCLab, 91405 Orsay, France}
\author{K. Hadyńska-Klęk}
\affiliation{Department of Physics, University of Oslo, Oslo, Norway}
\author{F. Ibrahim}
\affiliation{Université Paris-Saclay, CNRS/IN2P3, IJCLab, 91405 Orsay, France}
\author{N. Jovancevic}
\altaffiliation[Present address: ]{University of Novi Sad, Faculty of Science, Novi Sad, Serbia.}
\affiliation{Université Paris-Saclay, CNRS/IN2P3, IJCLab, 91405 Orsay, France}
\author{{A. Kankainen}}
\affiliation{Department of Physics, Accelerator Laboratory, P.O. Box 35, University of Jyväskylä, FI-40014 Finland}
\author{M. Lebois }
\affiliation{Université Paris-Saclay, CNRS/IN2P3, IJCLab, 91405 Orsay, France}
\author{{T. Mart\'{i}nez}\orcidlink{0000-0002-0683-5506}}
\affiliation{Centro de Investigaciones Energeticas Medioambientales y Tecnológicas (CIEMAT), Madrid, Spain}
\author{{P. Napiorkowski}}
\affiliation{Heavy Ion Laboratory, University of Warsaw, 02-093 Warsaw, Poland}
\author{B. Roussiere}
\affiliation{Université Paris-Saclay, CNRS/IN2P3, IJCLab, 91405 Orsay, France}
\author{{Yu. G. Sobolev}\orcidlink{0000-0002-5615-0468}}
\affiliation{Joint Institute for Nuclear Research, Dubna, Russia}
\author{{I. Stefan}\orcidlink{0000-0001-7923-8908}}
\affiliation{Université Paris-Saclay, CNRS/IN2P3, IJCLab, 91405 Orsay, France}
\author{{S. Stukalov}\orcidlink{0000-0003-2948-1947}}
\affiliation{Joint Institute for Nuclear Research, Dubna, Russia}
\author{D. Thisse}
\affiliation{Université Paris-Saclay, CNRS/IN2P3, IJCLab, 91405 Orsay, France}
\author{G. Tocabens}
\affiliation{Université Paris-Saclay, CNRS/IN2P3, IJCLab, 91405 Orsay, France}

\date{\today}

\begin{abstract}

The $\beta$-delayed spectroscopy of $^{80}$Ge has been studied using sources of ground and low-lying isomeric states of $^{80}$Ga. A hybrid $\gamma$-ray spectrometer was used, composed of high-purity germanium (HPGe) detectors for low-energy $\gamma$-ray detection, phoswich detectors from the PARIS array for high-energy $\gamma$ rays and a plastic detector for $\beta$ tagging. The new decay level schemes are presented, with 13 and 14 states $\beta$ populated by $^{80g}$Ga and $^{80m}$Ga, respectively, being reported for the first time. We quantitatively compare summed intensities of first-forbidden and previously reported Gamow-Teller $\beta$ transitions [R. Li et al., \href{https://doi.org/10.1103/PhysRevC.111.034303}{Phys. Rev. C 111, 034303 (2025)}]. The upper-limit fractions of first-forbidden transitions contributing to the decays of $^{80g}$Ga and $^{80m}$Ga are (48.0 $\pm$ 2.7)$\%$ and (47.9 $\pm$ 3.3)$\%$ of the observed total $\beta$ transition intensities of (78.2 $\pm$ 2.5)$\%$ and (82.2 $\pm$ 3.7)$\%$, respectively. Notably, the half-lives decrease in accordance with the upper limits of (48.0 $\pm$ 6.0)$\%$ [3.67(20) s $\rightarrow$ 1.91(3) s] and (47.9 $\pm$ 7.2)$\%$ [3.01(19) s $\rightarrow$ 1.57(1) s] for $^{80g}$Ga and $^{80m}$Ga, respectively, when including first-forbidden transitions, in contrast to those by Gamow-Teller transitions only. 

\end{abstract}

\maketitle

\section{I. INTRODUCTION}

$\beta$ decay of exotic nuclei offers valuable opportunities to explore weak interactions in the nuclear medium and to investigate the structure of the resulting daughter nuclei. In particular, the decay of $^{80}$Ga to $^{80}$Ge---a nucleus with 32 protons and 48 neutrons, corresponding to two neutron holes in the $\it{N}$ = 50 closed shell---presents several unique features that make it especially suitable for such studies. First, the $\beta$-decaying ground and isomeric states of $^{80}$Ga, with spin-parity values of 6$^-$ and 3$^-$, respectively, can populate a wide spin range of daughter states (from 0 to 8$\hbar$) in $^{80}$Ge, through both allowed and first-forbidden (FF) transitions. Second, the neutron-rich nature of $^{80}$Ga leads to a large Q$_{\beta}$ value of 10.312(4) MeV \cite{wang2021ame}, primarily due to its high isospin asymmetry (0.23). Third, while the structure of $^{80}$Ge is generally regarded as being governed by strong shell effects \cite{iwasaki2008persistence}, indications of quadrupole and triaxial deformations have also been proposed \cite{verney2013structure}. Consequently, $\beta$ decay of $^{80}$Ga allows to investigate how nuclear shell structure and collectivity influence $\beta$ decay properties in the vicinity of closed shells \cite{PhysRevC.111.034303}, and to probe specific structural features in $^{80}$Ge such as the pygmy dipole resonance (PDR) \cite{PhysRevC.110.064323}.

The first $\beta$ decay study of $^{80}$Ga was performed by Hoff and Fogelberg through neutron-induced fission of $^{235}$U, and the level scheme was determined up to 6.155 MeV. The energy region from 6.155 MeV to Q$_{\beta}$ value was left blank. The cumulated I$_{\beta}$ intensity was 73.94$\%$. Compared with $\%\beta^-\gamma$ (total branching ratios of $\beta$ decays followed by $\gamma$-ray emissions from the daughter nucleus) = 99.14, 25$\%$ branching-ratios were not observed \cite{hoff1981properties}. Following the observation of a low-lying isomeric state in $^{80}$Ga \cite{cheal2010discovery}, Verney $\it{et}$ $\it{al.}$ separated the decay level schemes of the ground state and isomeric state but only up to 3.5 MeV \cite{verney2013structure}. Recently, $\beta$ decay of a mixed source $^{80g+m}$Ga was performed using proton-induced fission with thick UC$_x$ target. However, this measurement suffered from serious contamination of $^{80}$Rb of 78$\%$ \cite{garcia2020absence}. Sekal $\it{et}$ $\it{al.}$ took advantage of the $^{80}$Zn decay chain to build a decay level scheme of $^{80m}$Ga only \cite{sekal2021low}. The latest study reported the Gamow-Teller (GT) strength distribution of $^{80}$Ga covering the entire $\beta$-decay window. This was achieved using a completely pure $^{80}$Ga beam produced in photofission of UC$_x$ induced by a 50 MeV electron beam, with 6 and 16 new states populated by $^{80g}$Ga and $^{80m}$Ga, respectively, and 56 newly observed $\gamma$ rays \cite{PhysRevC.111.034303}.

In this article, we disentangle the decay level schemes fed by the two $\beta$-decaying states of $^{80}$Ga in the whole Q$_{\beta}$ window. Separated decay level schemes up to S$_n$ are established. Using these new data, we investigate the competition between allowed (GT) and FF $\beta$ decays of $^{80}$Ga and study the impact of this competition on the lifetime of the precursor.

\section{II. EXPERIMENT}

The measurement was performed at the Acc{\'e}l{\'e}rateur Lin{\'e}aire et Tandem à Orsay (ALTO) ISOL facility \cite{ibrahim2007alto}. A radioactive low-energy $^{80}$Ga ion beam was produced by the photofission of a UC$_x$ target induced by a 50-MeV electron beam with an intensity of $\approx$7 $\mu$A. For beam purification, the laser ionized Ga beam was mass selected. Since at around $\it{A}$ = 80 the only surface ionized component of a photofission generated ion beam is Ga, complete isotope purity was achieved, with zero $^{80}$Rb contamination. This cleanness was tested by the $\beta$-gated $\gamma$ spectrum. No contaminated $\gamma$ rays, like the $\beta^+$/$\epsilon$-delayed 616.7-keV $\gamma$ line of $^{80}$Kr, were found in the spectrum. Only $\gamma$ rays from $^{80}$Ge and its daughter and granddaughter nuclei were observed. Then, a $^{80}$Ga beam with yield of $\approx$10$^{4}$ pps was directed and collected by the tape system of the BEDO (BEta Decay studies at Orsay) setup, which was moved periodically for minimizing the daughters activity \cite{etile2015low}. The time settings were 0.5 s for background, 5 s for ion collection, and 5 s for decay measurement.

The emitted radiations were detected by a hybrid array around the collection spot. It consisted of one cylindrical plastic scintillator for $\beta$ tagging, surrounded by two high-purity germanium detectors (HPGe), one clover HPGe and one coaxial HPGe, and three PARIS clusters \cite{ciemala2009measurements,ghosh2016characterization}. Each cluster comprises nine optically isolated segmented phoswiches. A phoswich is composed of two layers of different scintillators and a photomultiplier. The first layer is a 2 in.$\hspace{0.16cm}\times$2 in.$\hspace{0.16cm}\times$2 in.$\hspace{0.16cm}$lanthanum bromide crystal (LaBr$_3$); one of the three clusters is cerium bromide (CeBr$_3$). The second layer is a 2 in. $\times$2 in.$\hspace{0.15cm}\times$6 in.$\hspace{0.15cm}$sodium iodide (NaI) crystal.$\hspace{0.2cm}$The high energy resolution of the HPGe working in the  0$-$6 MeV energy range makes very effective the  $\gamma$-$\gamma$ coincidence technique not only to reconstruct the transition cascades but also to suppress the background drastically. PARIS working in the 6$-$10 MeV energy range has high detection efficiency. The detectors were energy calibrated up to 9 MeV using sources including $^{58}$Ni(n$_{th}$,$\gamma$) with energy of 8999.267(15) keV. The energy resolution of HPGe detectors is 2.5 keV at 1109 keV (a $\gamma$ line of $^{80}$Ge) from summed spectrum. Considering 10.39(2), 12.65(5), 1.65(2), and 2.78(4) ns time resolutions of coaxial HPGe, clover HPGe, LaBr$_3$(Ce), and NaI, respectively, 50 ns was chose as the $\gamma$-$\gamma$ coincidence time window. The detection efficiency of HPGe detectors including one clover HPGe and one coaxial HPGe is 3.134(5)$\%$ at 1085.8 keV. The energy resolution and detection efficiency of PARIS clusters are 112 keV and 0.30(2)$\%$, respectively, at 7.18 MeV, under mode 3 \cite{PhysRevC.111.034303}; signals only come from LaBr$_3$(Ce) crystals and are vetoed by outer-layer NaI detectors, but the energy is added back within 27 phoswich detectors. Data were acquired in a triggerless mode.

For extracting precise log $\it{ft}$ and $\it{B}$(GT) values, counting precursors with high precision is critical. The applied methodology was handling the Bateman equations using the total $\beta$ spectrum. This allows one to determine the activity of the precursors and the intensity of the radioactive beam. The final counts of precursor $^{80}$Ga are N$_{^{80m}Ga}^{decayed}$ = 1.41(8)$\times10^8$ and N$_{^{80g}Ga}^{decayed}$ = 2.40(8)$\times10^8$. The uncertainties on these numbers originate from the uncertainties of input parameters including half-lives of precursor and daughter and granddaughter nuclei, P$_n$, and the detection efficiency of the $\beta$ detector.

In order to obtain two separated decay level schemes of $^{80g}$Ga and $^{80m}$Ga, it is essential to identify the $\beta$ feeding precursor for each state of the daughter nucleus. For the identification methods, we refer to Refs. \cite{verney2013structure,PhysRevC.111.034303}. 

\section{III. EXPERIMENTAL RESULTS}

\begin{figure*}
    \centering
    \includegraphics[width=1.0\textwidth]{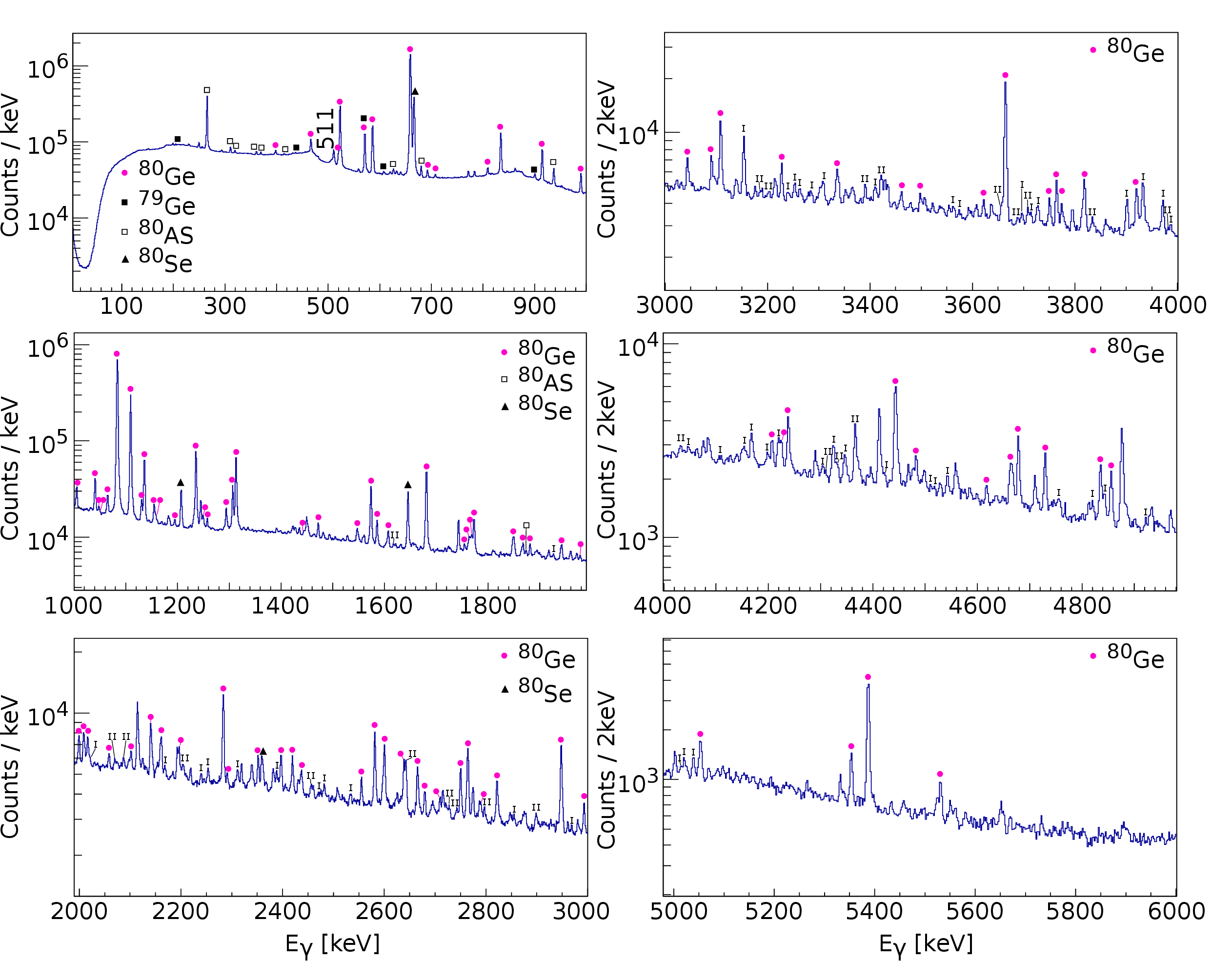}
    \caption{$\beta$-gated $\gamma$ spectra in the range 0 to 6000 keV. The marked lines are attributed to transitions in daughter nuclei: filled circles = $^{80}$Ge; filled squares = $^{79}$Ge; empty squares = $^{80}$As and filled triangles = $^{80}$Se. I and II symbols denote single- and double-escape peaks, respectively. Note that I and II are not sequential.}
    \label{fig6.30}
\end{figure*}

\begin{figure*}
    \centering
    \includegraphics[width=0.9\textwidth]{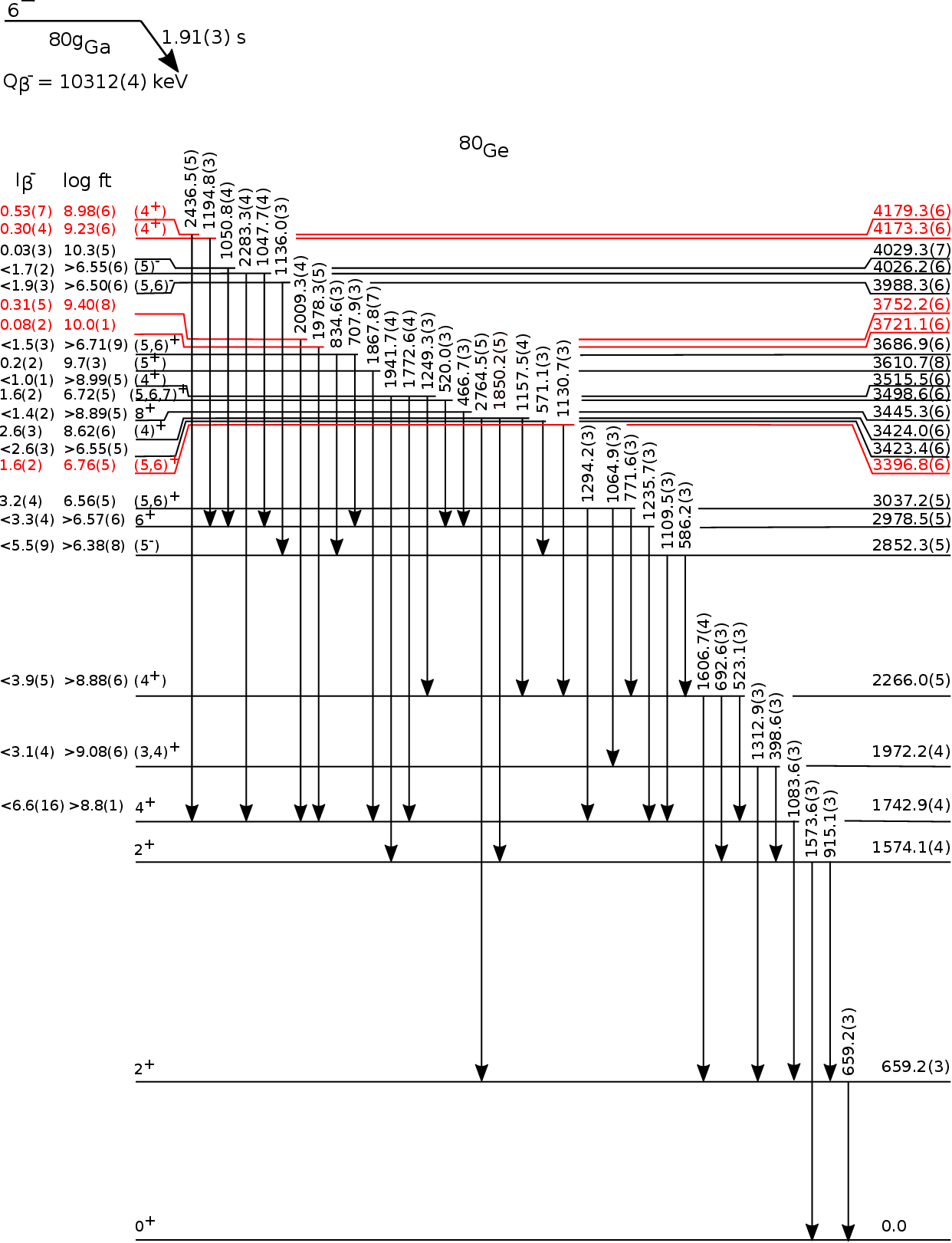}
    \caption{Level scheme of $^{80}$Ge up to 4.2 MeV in excitation energy populated following the $\beta$ decay of $^{80g}$Ga. For the sake of clarity, the decay scheme has been split in two sections, with the one for the higher energies plotted in Fig. \ref{fig6.41}.}
    \label{fig6.40}
\end{figure*}

\begin{figure*}
    \centering
    \includegraphics[width=0.9\textwidth]{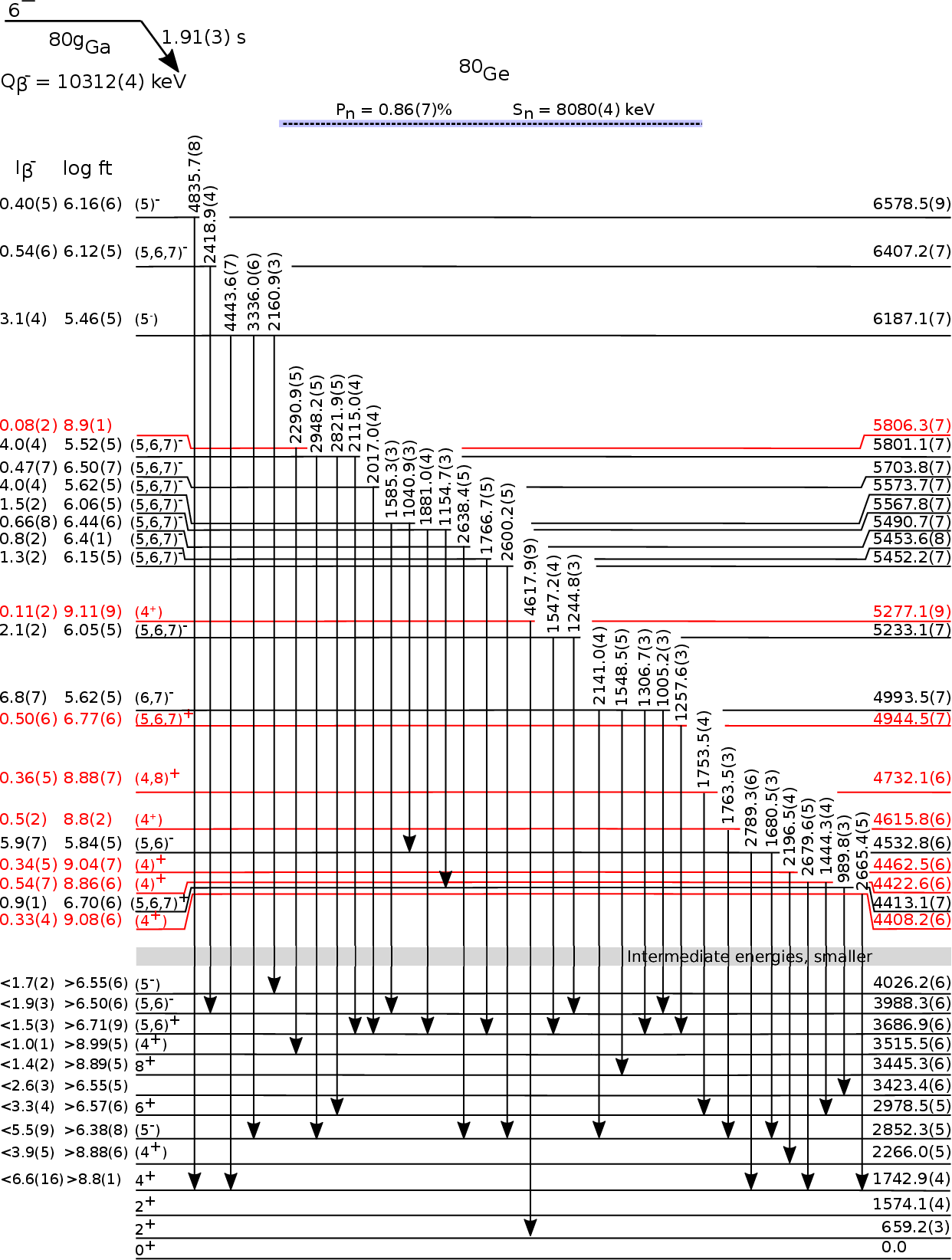}
    \caption{Level scheme of $^{80}$Ge populated following the $\beta$ decay of $^{80g}$Ga containing the high-lying states between 4.2 and 8 MeV in excitation energy.}
    \label{fig6.41}
\end{figure*}

The $\beta$-gated $\gamma$ spectra measured during the experiment are presented in Fig. \ref{fig6.30}. One can observe the $\beta$-delayed $\gamma$ rays belonging to daughter and granddaughter nuclei $^{80}$Ge, $^{80}$As, and $^{80}$Se with different marks. In addition, $\gamma$-ray deexcitation of the nucleus $^{79}$Ge, populated through $\beta$-delayed neutron emission of $^{80}$Ga, is also detected. No other $\gamma$ rays were observed in the spectra, which proves the beam purity. All these $\gamma$ rays populated through deexcitation of the excited states of $^{80}$Ge allowed us to improve and enrich the decay level scheme of $^{80}$Ge. In the present work, a total of 112 characteristic half-lives of activities of $\beta$-delayed $\gamma$ rays could be determined; see tabulated $\gamma$ information in Table I in Ref. \cite{PhysRevC.111.034303}. Among them 70 were assigned to the decay of $^{80g}$Ga and 65 to the decay of $^{80m}$Ga. Twenty-three $\gamma$ rays are emitted by both isomers. Among the 70 $\gamma$ rays, 29 $\gamma$ rays are observed for the first time, indicated by the red color lines in the decay level scheme in Figs. \ref{fig6.40} and \ref{fig6.41}. Among the 65 $\gamma$ rays, 35 $\gamma$ rays are detected and reported for the first time. 

\begin{figure*}
    \centering
    \includegraphics[width=0.9\textwidth]{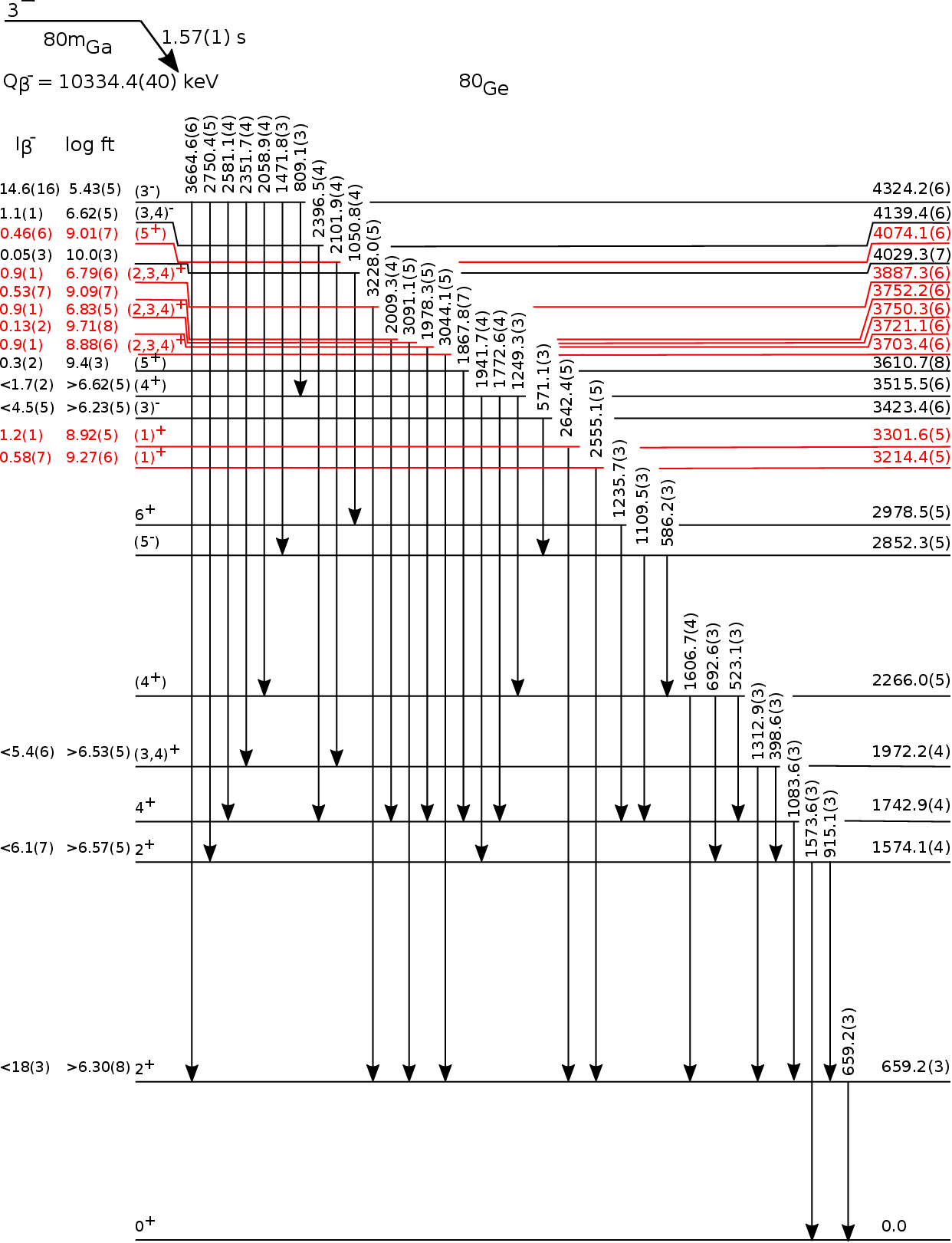}
    \caption{Level scheme of $^{80}$Ge up to 4.4 MeV in excitation energy populated following the $\beta$ decay of $^{80m}$Ga. For the sake of clarity, the decay scheme has been split in two sections, with the one for the higher energies plotted in Fig. \ref{fig6.43}.}
    \label{fig6.42}
\end{figure*}

\begin{figure*}
    \centering
    \includegraphics[width=0.9\textwidth]{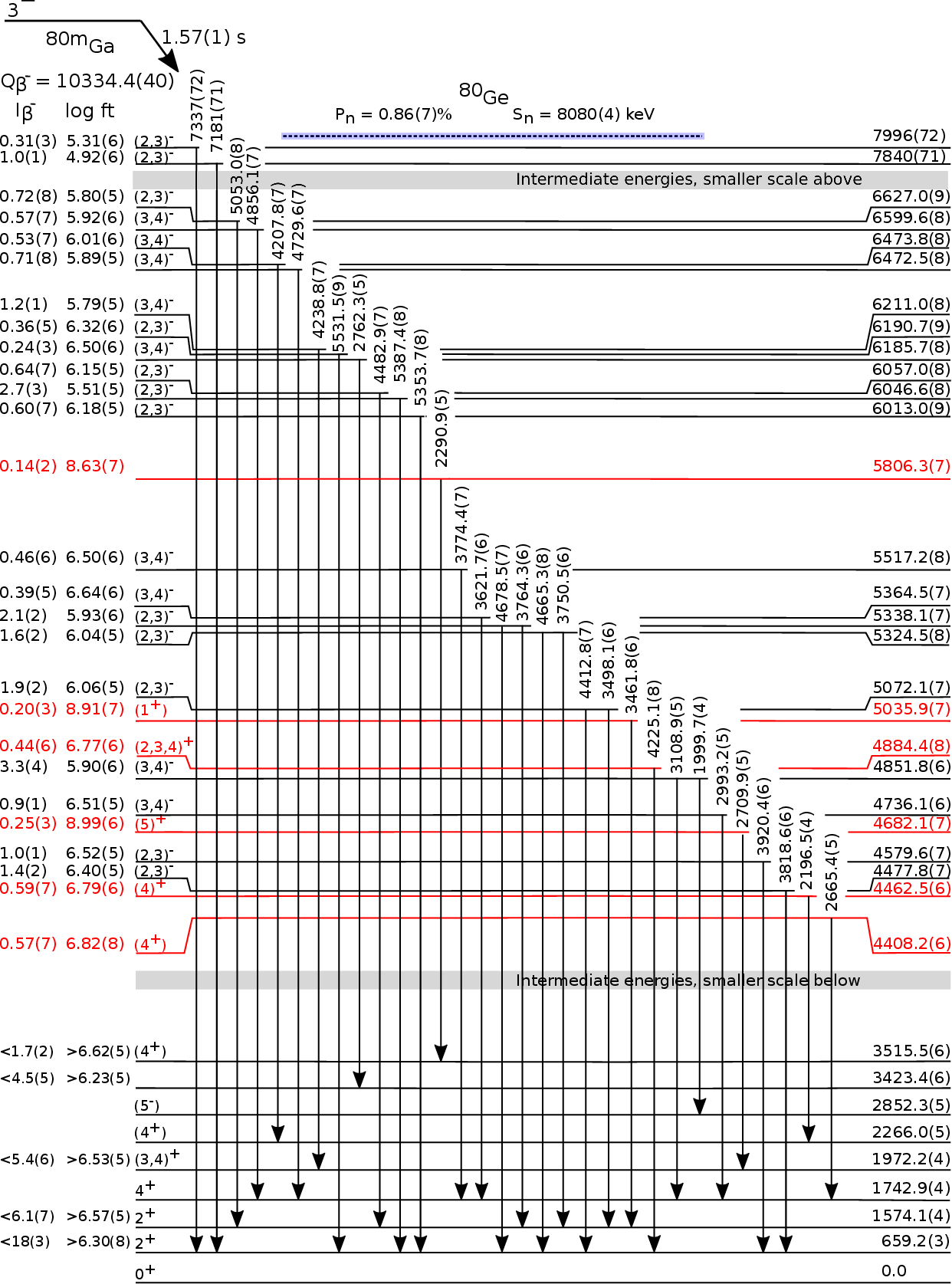}
    \caption{Level scheme of $^{80}$Ge populated following the $\beta$ decay of $^{80m}$Ga containing the high-lying states between 4.4 and 8 MeV in excitation energy.}
    \label{fig6.43}
\end{figure*}

For the precursor identification, only ten states out of a total of 77 populated states were assigned to $\beta$ feedings from both $^{80g}$Ga and $^{80m}$Ga simultaneously. They are the levels with the energies of 1972.2, 3423.4, 3515.5, 3610.7, 3721.1, 3752.2, 4029.3, 4408.2, 4462.5, and 5806.3 keV. The summed I$_{\beta}$ of these ten states are 8.1(6)$\%$ and 13.9(8)$\%$ in the decay level schemes of $^{80g}$Ga and $^{80m}$Ga, respectively. The $\gamma$ lines from these states are assigned as 50$\%$ from $^{80g}$Ga and 50$\%$ from $^{80m}$Ga. It is worth pointing out that having the intensity distributed in both level schemes for some $\gamma$ rays and levels can introduce a systematic error in the feedings. Further checks of these ten states with high statistics deserve being pursued in future experiments.

From the present data, one can place 45 excited states of $^{80}$Ge populated in the $\beta$ decay of $^{80g}$Ga as shown in Figs. \ref{fig6.40} and \ref{fig6.41}. Thirteen states are reported for the first time. The scheme was built based on the $\beta$-$\gamma$-$\gamma$ coincidence relationships. The excitation energies of states were obtained through calculating the weighted average value of distinct deexcitation cascades. For example, the excitation energy for the state at 5801.1(7) keV was obtained by considering the three following deexcitation cascades: 2948.2(5) + 2852.3(5) keV, 2821.9(5) + 2978.5(5) keV, and 2115.0(4) + 3686.9(6) keV. The weight is according to the strength (I$_{\gamma}^{rel}$) of the three $\gamma$ rays 2948.2(5), 2821.9(5) and 2115.0(4) keV. They are 2.42(9)$\%$, 1.01(6)$\%$ and 2.39(9)$\%$ individually. Then, a weighted energy level of 5801.1 keV state is obtained. The uncertainty 0.7 keV is also a weighted result of the three cascades.

The I$_{\beta}$ values were calculated using the balance of the observed feeding and depopulating $\gamma$ activities, which means that, for a given state, the $\beta$ feeding is the difference between the $\gamma$ feeding to and $\gamma$ decay from this state. Since the cumulated values of I$_{\beta}$ from $^{80g}$Ga and $^{80m}$Ga reach 79.1(25)$\%$ and 85.3(36)$\%$, respectively, one can conclude that some states are unobserved in this work, especially in the high energy region around S$_n$ (8.08 MeV). These states would deexcite to some low-lying states like 1742.9, 2266.0, 2852.3, 2978.5, 3423.4, 3445.3, 3515.5, 3686.9, 3988.3, and 4026.2 keV. In order to avoid being overweighted, the I$_{\beta}$ of these states were just given an upper limit in this work.

The log $\it{ft}$ value calculations were performed using the NNDC (National Nuclear Data Center) online procedure \cite{logft}, wherein $\it{T}_{1/2}$, E$_{energy}$, and I$_{\beta}$ were from this work and the used Q$_{\beta}$ was taken from Ref. \cite{wang2021ame}.

As presented in Figs. \ref{fig6.42} and \ref{fig6.43}, 48 of excited states of $^{80}$Ge were populated by $^{80m}$Ga $\beta$ decay and 14 of these states are proposed for the first time (shown with red colors) in the current work. Of the 14 new states, 6 are located interestingly above 4 MeV. This means the I$_{\beta}$ values of low-lying states were overweighted in previous work while they were underweighted in the high-energy region. This was caused by unobserved high-energy $\gamma$ rays. The states at 7840(71) and 7996(72) keV were determined based on the observation of high-energy $\gamma$ rays, 7181(71) and 7337(72) keV, by the PARIS array. I$_{\beta}$, log $\it{ft}$ and their uncertainty analysis procedures are as stated before. It is worth noting that $\it{Q}_{\beta}$ for $^{80m}$Ga is 10334.4(40) keV, including 22.4 keV \cite{wang2021ame} corresponding to the excitation energy of the isomeric state. The spin-parity assignment will be discussed in the next section.

\section{IV. DISCUSSION}

log $\it{ft}$ values are among the most important observables in $\beta$ decay studies, and which are affected by the nuclear structure of the parent and daughter nuclei. They are also vital experimental data that can be used to support theory development. We systematically analyzed the log $\it{ft}$ values of $\it{N}$ = 49 isotones; see Table 5.5 in Ref. \cite{ren2022thesis}; it provides a reference to determine the types of $\beta$ transitions. Along with the proton number increasing, the log $\it{ft}$ values of allowed and FF transitions also increase. This means that in the $\it{N}$ = 49 isotonic chain from exotic to stable nuclei the allowed and FF $\beta$ decay probabilities decrease.

In this neutron-rich region \cite{NOWACKI2021103866}, it is often difficult to separate allowed from FF transitions since the latter become significant for neutron-rich nuclei. The references for the identification of $\beta$ transition type are global statistics for the whole nuclear chart and the local standard. In the prior reviews \cite{singh1998review,TURKAT2023101584}, the log $\it{ft}$ values are 3$-$4 for superallowed $\beta$ decay, 4$-$8 for allowed transition, 6$-$9 for FF transition, and 8$-$11 for first-forbidden unique (1U) transition. Since the 0$_1^+$, 2$_1^+$, 4$_1^+$, 6$_1^+$ and 8$_1^+$ levels are well identified in $^{80}$Ge, the transitions 6$^-$($^{80g}$Ga)$\rightarrow$4$_1^+$, 6$^-$($^{80g}$Ga)$\rightarrow$6$_1^+$ and 6$^-$($^{80g}$Ga)$\rightarrow$8$_1^+$ can provide a local reference of FF and 1U transitions. Likewise, 3$^-$($^{80m}$Ga)$\rightarrow$2$_1^+$ and 3$^-$($^{80m}$Ga)$\rightarrow$2$_2^+$ provide references for FF transition as well.

Based on this analysis, among the $\beta$ transitions of $^{80g}$Ga, the allowed transition character was assigned when log $\it{ft}$ was smaller than 6.6, which populated (5,6,7)$^-$ states; FF transition was assigned when log $\it{ft}$ was larger than 6.6 and smaller than 6.8, which populated (5,6,7)$^+$ states; 1U transition was assigned when log $\it{ft}$ was larger 6.8, and the log $\it{ft}$ value was calculated again using the 1U formula, which populated (4,8)$^+$. For $^{80m}$Ga, the same identifying standards were adopted. Therefore, allowed transitions populated (2,3,4)$^-$ states; FF transitions populated (2,3,4)$^+$ states; 1U transitions populated (1,5)$^+$ states. This analysis is based on the hypothesis that the spin-parity values of $^{80g+m}$Ga (6$^-$ and 3$^-$) are correct \cite{cheal2010discovery}, and that the second forbidden transition with "log $\it{ft}$ > 11" in the global statistics \cite{singh1998review,TURKAT2023101584} was negligible. Combined with this $\beta$ transition identification, the candidate spin-parity of states of $^{80}$Ge were given on the prior consideration of $\it{E}$1, $\it{E}$2 and $\it{M}$1 multipolarities of $\gamma$ deexcitations.

The final results are presented in Tables \ref{tab1} and \ref{tab2}. A total of 43 $\beta$ decays from the 6$^-$ state and 44 from the 3$^-$ state were determined. The existence of unobserved $\gamma$ rays, for low-lying states, could significantly decrease the deduced $\beta$ feedings and consequently increase the log $\it{ft}$ values. Therefore, the log $\it{ft}$ values of low-lying states in the $^{80g+m}$Ga decay level schemes were just provided as minimum values, which guarantees that the branching ratios of low-lying states are not overestimated. Besides I$_{\beta}$, log $\it{ft}$ and J$^{\pi}$ information, the $\it{X}$ value of each state is also presented in the tables. $\it{X}$ is the fraction of the direct $\beta$ feeding from the 3$^-$ isomer, which includes these different contributions and should take a value 0 (1) if the level is populated by $\beta$ decay of $^{80g}$Ga ($^{80m}$Ga). For more details about the definition of $\it{X}$ value and its usefulness in precursor identification, we refer to Refs. \cite{verney2013structure,PhysRevC.111.034303}.

\begin{table}[htbp] 
\centering
\caption{Excited states of $^{80}$Ge $\beta$ fed by $^{80g}$Ga.}  \vspace{0.5mm}
\label{tab1}
\renewcommand\arraystretch{0.7}
\begin{tabular}{lllll}
$E_{state}$ (keV) \hspace{0.2 in}	& $I_{\beta}$ \hspace{0.2 in}		& log $\it{ft}$ \hspace{0.2 in}	& $J^{\pi}$ \hspace{0.2 in}	 & $\it{X}$		\\
\hline
\hline
\colrule
1742.9(4) \hspace{0.2 in}		& <6.6(16) \hspace{0.2 in}		& >8.8(1) \hspace{0.2 in}		& 4$^{+}$ \hspace{0.2 in}	& 0.1(14)		\\
1972.9(4)		& <3.1(4)		& >9.08(6)		& (3,4)$^{+}$	& 0.60(32)			\\
2266.0(5)		& <3.9(5)		& >8.88(6)		& (4$^{+}$)	& 0.33(95)	\\
2852.3(5)		& <5.5(9)		& >6.38(8)		& (5$^{-}$)	& $-$0.2(12)	\\
2978.5(5)		& <3.3(4)		& >6.57(6)		& 6$^{+}$	& $-$0.37(34)	\\
3037.2(5)		& 3.2(4)		& 6.56(5)		& (5,6)$^{+}$	& 0.25(16)		\\
3396.8(6)		& 1.6(2)		& 6.76(5)		& (5,6)$^{+}$	& 0.31(25)	\\
3423.4(6)		& <2.6(3)		& >6.55(5)		& 				& 0.56(32)	\\
3424.0(6)		& 2.6(3)		& 8.62(6)		& (4$^{+}$)		& 0.40(31)	\\
3445.3(6)		& <1.4(2)		& >8.89(5)		& 8$^{+}$	& 0.10(34)		\\
3498.6(6)		& 1.6(2)		& 6.72(5)		& (5,6,7)$^{+}$	& 0.17(24)	\\
3515.5(6)		& <1.0(1)		& 8.99(5)		& (4$^{+}$)	& 0.55(42)		\\
3610.7(8)		& 0.2(2)		& 9.7(3)		& (5$^{+}$)	& 0.5(59)	\\
3686.9(6)		& <1.5(3)		& >6.71(9)		& (5,6)$^{+}$	& $-$0.7(17)		\\
3721.1(6)		& 0.08(2)		& 10.0(1)		& 			& 0.5(14)	\\
3752.2(6)		& 0.31(5)		& 9.40(8)		& 			& 0.42(70)	\\
3988.3(6)		& <1.9(3)		& >6.50(6)		& (5,6)$^{-}$	& $-$0.33(74)	\\
4026.2(6)		& <1.7(2)		& >6.55(6)		& (5$^{-}$)	& 0.16(66)	\\
4029.3(7)		& 0.03(3)		& 10.3(5)		& 			& 0.4(59)	\\
4173.3(6)		& 0.30(4)		& 9.23(6)		& (4$^{+}$)	& 0.32(75)	\\
4179.3(6)		& 0.53(7)		& 8.98(6)		& (4$^{+}$)	& 0.22(67)	\\
4408.2(6)		& 0.33(4)		& 9.08(6)		& (4$^{+}$)	& 0.54(44)	\\
4413.1(7)		& 0.9(1)		& 6.70(6)		& (5,6,7)$^{+}$	& 0.05(89)	\\
4422.6(6)		& 0.54(7)		& 8.86(6)		& (4$^{+}$)	& 0.14(54)	\\
4462.5(6)		& 0.34(5)		& 9.04(7)		& (4)$^{+}$	& 0.49(63)	\\
4532.8(6)		& 5.9(7)		& 5.84(5)		& (5,6)$^{-}$	& 0.02(22)		\\
4615.8(6)		& 0.5(2)		& 8.8(2)		& (4$^{+}$)	& 0.3(41)	\\
4732.1(6)		& 0.36(5)		& 8.88(7)		& (4,8)$^{+}$	& 0.25(67)		\\
4944.5(7)		& 0.50(6)		& 6.77(6)		& (5,6,7)$^{+}$	& 0.24(42)	\\
4993.5(7) 		& 6.8(7)		& 5.62(5)		& (6,7)$^{-}$	& 0.12(14)		\\
5233.1(7)		& 2.1(2)		& 6.05(5)		& (5,6,7)$^{-}$	& 0.18(20)	\\
5277.1(9)		& 0.11(2)		& 9.11(9)		& (4$^{+}$)		& $-$0.1(20)	\\
5452.2(7)		& 1.3(2)		& 6.15(5)		& (5,6,7)$^{-}$	& 0.18(25)	\\
5453.6(8)		& 0.8(2)		& 6.4(1)		& (5,6,7)$^{-}$	& 0.1(26)	\\
5490.7(7)		& 0.66(8)		& 6.44(6)		& (5,6,7)$^{-}$	& 0.19(48)	\\
5567.8(7)		& 1.5(2) 		& 6.06(5)		& (5,6,7)$^{-}$	& 0.20(39)	\\
5573.7(7)		& 4.0(4)		& 5.62(5)		& (5,6,7)$^{-}$	& 0.11(15)	\\
5703.8(7)		& 0.47(7)		& 6.50(7)		& (5,6,7)$^{-}$	& 0.24(95)	\\
5801.1(7)		& 4.0(4)		& 5.52(5)		& (5,6,7)$^{-}$	& 0.22(13)	\\
5806.3(7)		& 0.08(2)		& 8.9(1)		& 				& 0.6(12)	\\
6187.1(7)		& 3.1(4)		& 5.46(5)	    & (5$^{-}$)		& 0.13(32)	\\
6407.2(7)		& 0.54(6)		& 6.12(5)		& (5,6,7)$^{-}$	& 0.01(45)	\\
6578.5(9)		& 0.40(5)		& 6.16(6)		& (5)$^{-}$		& $-$0.16(71)	\\
\hline
\hline

\end{tabular}
\end{table}

\begin{table}[htbp] 
\centering
\caption{Excited states of $^{80}$Ge $\beta$ fed by $^{80m}$Ga.}  \vspace{0.5mm}
\label{tab2}
\renewcommand\arraystretch{0.7}
\begin{tabular}{lllll}
$E_{state}$ (keV) \hspace{0.2 in}	& $I_{\beta}$ \hspace{0.2 in}	& log $\it{ft}$ \hspace{0.2 in}		& $J^{\pi}$ \hspace{0.2 in}	 & $\it{X}$	\\
\hline
\hline
659.2(3) \hspace{0.2 in}		& <18(3) \hspace{0.2 in}		& >6.30(8) \hspace{0.2 in}	& 2$^{+}$ \hspace{0.2 in} 		& 1.0(20)	\\
1574.1(4)		& <6.1(7)		& >6.57(5)		& 2$^{+}$	& 0.83(27)	\\
1972.2(4)		& <5.4(6)		& >6.53(5)		& (3,4)$^{+}$	& 0.60(32)	\\
3214.4(5)		& 0.58(7)		& 9.27(6)		& (1)$^{+}$	& 0.82(72)	\\
3301.6(5)		& 1.2(1)		& 8.92(5)		& (1)$^{+}$	& 0.46(30)	\\
3423.4(6)		& <4.5(5)		& >6.23(5)		& (3)$^-$	& 0.56(32)	\\
3515.5(6)		& <1.7(2)		& >6.62(5)		& (4$^{+}$)	& 0.55(42)	\\
3610.7(8)		& 0.3(2)		& 9.4(3)		& (5$^{+}$)	& 0.5(59)	\\
3703.4(6)		& 0.9(1)		& 8.88(6)		& (2,3,4)$^{+}$	& 0.83(69)		\\
3721.1(6)		& 0.13(2)		& 9.71(8)		& 				& 0.5(14)	\\
3750.3(6)		& 0.9(1)		& 6.83(5)		& (2,3,4)$^{+}$	& 0.65(55)	\\
3752.2(6)		& 0.53(7)		& 9.09(7)		& 				& 0.42(70)	\\
3887.3(6)		& 0.9(1)		& 6.79(6)		& (2,3,4)$^{+}$	& 0.77(65)	\\
4029.3(7)		& 0.05(3)		& 10.0(3)		& 			& 0.4(59)	\\
4074.1(6)		& 0.46(6)		& 9.01(7)		& (5$^{+}$)	& 1.3(12)	\\
4139.4(6)		& 1.1(1)		& 6.62(5)		& (3,4)$^{-}$	& 0.67(57)	\\
4324.2(6)		& 14.6(16)		& 5.43(5)		& (3$^{-}$)	& 0.72(12)	\\
4408.2(6)		& 0.57(7)		& 6.82(8)		& (4$^{+}$)	& 0.54(44)	\\
4462.5(6)		& 0.59(7)		& 6.79(6)		& (4)$^{+}$	& 0.49(63)	\\
4477.8(7)		& 1.4(2)		& 6.40(5)		& (2,3)$^{-}$	& 0.85(74)	\\
4579.6(7)		& 1.0(1)		& 6.52(5)		& (2,3)$^{-}$	& 1.17(67)	\\
4682.1(7)		& 0.25(3)		& 8.99(6)		& (5$^{+}$)		& 0.8(18)	\\
4736.1(6)		& 0.9(1)		& 6.51(5)		& (3,4)$^{-}$	& 0.65(61)	\\
4851.8(6)		& 3.3(4)		& 5.90(6)		& (3,4)$^{-}$	& 0.60(22)	\\
4884.4(8)		& 0.44(6)		& 6.77(6)		& (2,3,4)$^{+}$	& 1.2(19)	\\
5035.9(7)		& 0.20(3)		& 8.91(7)		& (1$^{+}$)		& 1.2(22)	\\
5072.1(7)		& 1.9(2)		& 6.06(5)		& (2,3)$^{-}$	& 0.84(41)	\\
5324.5(8)		& 1.6(2)		& 6.04(5)		& (2,3)$^{-}$	& 0.76(58)	\\
5338.1(7) 		& 2.1(2)		& 5.93(6)		& (2,3)$^{-}$	& 0.60(46)	\\
5364.5(7)		& 0.39(5)		& 6.64(6)		& (3,4)$^{-}$	& 0.88(14)	\\
5517.2(8)		& 0.46(6)		& 6.50(6)		& (3,4)$^{-}$	& 1.2(17)	\\
5806.3(7)		& 0.14(2)		& 8.63(7)		& 				& 0.6(12)	\\
6013.0(9)		& 0.60(7)		& 6.18(5)		& (2,3)$^{-}$	& 1.22(84)	\\
6046.6(8)		& 2.7(3)		& 5.51(5)		& (2,3)$^{-}$	& 1.19(32)	\\
6057.0(8)		& 0.64(7)		& 6.15(5)		& (2,3)$^{-}$	& 0.59(58)	\\
6185.7(8)		& 0.24(3) 		& 6.50(6)		& (3,4)$^{-}$	& 0.7(16)	\\
6190.7(9)		& 0.36(5)		& 6.32(6)		& (2,3)$^{-}$	& 0.7(13)	\\
6211.0(8)		& 1.2(1)		& 5.79(5)		& (3,4)$^{-}$	& 1.02(64)	\\
6472.5(8)		& 0.71(8)		& 5.89(5)		& (3,4)$^{-}$	& 1.45(85)	\\
6473.8(8)		& 0.53(7)		& 6.01(6)		& (3,4)$^{-}$	& 0.7(13)	\\
6599.6(8)		& 0.57(7)		& 5.92(6)    	& (3,4)$^{-}$	& 0.80(67)	\\
6627.0(9)		& 0.72(8)		& 5.80(5)	    & (2,3)$^{-}$	& 0.78(79)	\\
7840(71)		& 1.0(1)		& 4.92(6)		& (2,3)$^{-}$	& 0.83(33)	\\
7996(72)		& 0.31(3)		& 5.31(6)		& (2,3)$^{-}$	& 1.24(56)	\\
\hline
\hline

\end{tabular}
\end{table}

The spin-parity values of states populated by $\beta$ transitions with reduced log $\it{ft}$ values less than 6.0 can be adopted with high confidence. In other words, these transitions can be assigned to Gamow-Teller transitions. For the states with $\beta$ transition log $\it{ft}$ values between 6 and 6.6, their spin-parity assignments are regarded as tentative due to the serious competition between allowed and FF $\beta$ transitions. Furthermore, Fermi decay would populate 3$^-$ and 6$^-$ states in $^{80}$Ge, but such transitions only happen with low log $\it{ft}$ values to isobaric analog states (IAS), which are located at much higher energy than levels observed in this study. In addition, isospin mixing in the low-lying states (below S$_n$ = 8.08 MeV) is expected to be extremely small, which is reflected in the log $\it{ft}$ values, as one can find that there is no state which has a log $\it{ft}$ value of $\approx$3.5.

\begin{figure*}
    \centering
    \includegraphics[width=1.0\textwidth]{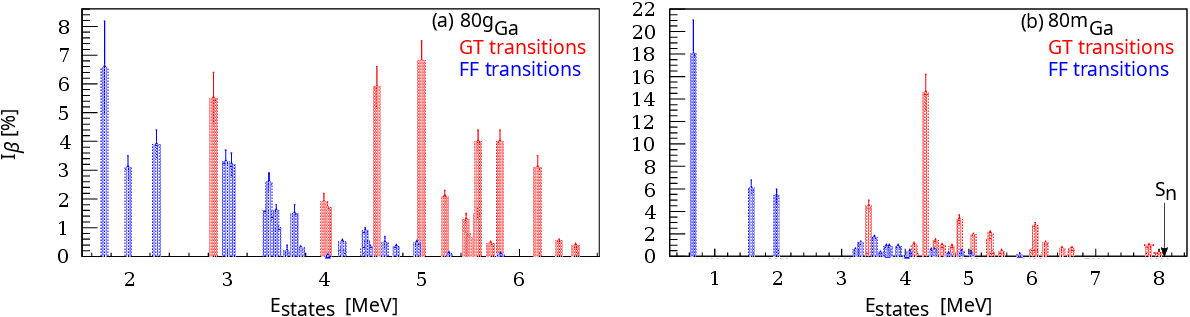}
    \caption{GT and FF $\beta$-population intensities as functions of excitation energy: (a) for $^{80g}$Ga and (b) for $^{80m}$Ga. Percentages of FF transitions contributing to the decays of $^{80g}$Ga and $^{80m}$Ga are (48.0 $\pm$ 2.7)$\%$ and (47.9 $\pm$ 3.3)$\%$, respectively.}
    \label{fig7.0}
\end{figure*}

Based on these new data, we study the competition between GT and FF transitions in $^{80g+m}$Ga $\beta$ decays, and the influence of this competition on the half-life of the precursor. Half-lives of neutron-rich nuclei set the time scale of the $\it{r}$ process, a postulated rapid-neutron-capture process happening in hot and neutron-rich stellar environments to synthesize heavy nuclei beyond Fe \citep{PhysRevC.67.055802,PhysRevC.67.025802,BORZOV2006645,PhysRevLett.113.022702,PhysRevC.93.025805,PhysRevLett.125.192501,PhysRevC.103.054327}; e.g., FF transition shortens the half-life of $\beta$ decay and shifts the $\it{r}$-abundance peak towards higher masses \cite{PhysRevC.85.015802,NISHIMURA2016273}. Due to collective excitations including vibrational and rotational motions, the space of the low-energy region of the even-even nucleus $^{80}$Ge is filled by positive-parity states such as (2-4-6-8)$_1^+$, (2-4-6-8)$_2^+$, and (3-5-7)$_1^+$. $\beta$ feedings from $^{80g}$Ga (6$^-$) and $^{80m}$Ga (3$^-$) to these low-lying states are FF transitions. Compared with $\beta$ feedings to high-lying states, these transitions have larger energy windows. According to the Sargent rule \citep{sargent1933maximum}, the decay constant $\lambda$ is proportional to the fifth power of the maximum kinetic energy of the $\beta$ particle ($\lambda$ $\propto$ $\it{Q}^5$). Hence, the probabilities of these transitions are no longer small, though they are forbidden transitions, i.e., prohibitions are "softened" due to phase-space amplification. Large I$_{\beta}$ values in blue colors located in the low-energy regions in Fig. \ref{fig7.0} illustrate this point. On the other hand, as the excitation energy increases, $\beta$ transitions reach the tail of the Gamow-Teller giant resonance (GTGR). Therefore, GT transitions become dominant. Furthermore, the level density in the high excitation-energy region is much larger than in the low excitation-energy region, i.e., there are more $\beta$ transitions possible to the high excitation-energy region. Dense and strong GT transitions marked with red colors located in the high excitation-energy regions in Fig. \ref{fig7.0} prove this point. For these reasons, a competition between FF and GT $\beta$ transitions arises in the neutron-rich nucleus $^{80}$Ga with a large value of isospin. The former shortens while the latter lengthens the half-life of the precursor.

\begin{figure}
    \centering
    \includegraphics[width=1.0\columnwidth]{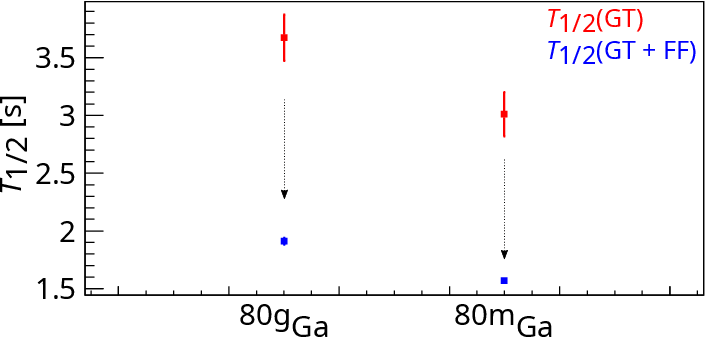}
    \caption{\justifying {Comparisons between partial half-lives of summed GT transitions (red squares) and half-lives of total transitions (blue squares). When including FF transitions, the half-lives decrease by (48.0 $\pm$ 6.0)$\%$ [3.67(20) s $\rightarrow$ 1.91(3) s] and (47.9$\pm$7.2)$\%$ [3.01(19) s $\rightarrow$ 1.57(1) s] for $^{80g}$Ga and $^{80m}$Ga, respectively.}}
    \label{fig7.1}
\end{figure}

From experimental results in Tables \ref{tab1} and \ref{tab2}, the upper-limit fractions of FF transitions contributing to the decays of $^{80g}$Ga and $^{80m}$Ga are (48.0 $\pm$ 2.7)$\%$ and (47.9 $\pm$ 3.3)$\%$ of the observed total $\beta$ transition intensities of (78.2 $\pm$ 2.5)$\%$ and (82.2 $\pm$ 3.7)$\%$, respectively, as shown in Fig. \ref{fig7.0}. The half-lives decrease in accordance with the upper limits of (48.0 $\pm$ 6.0)$\%$ and (47.9$\pm$7.2)$\%$ for $^{80g}$Ga and $^{80m}$Ga, respectively, when including FF transitions [$\it{T}_{1/2}$(GT+FF)] in contrast to those due to Gamow-Teller transitions only [$\it{T}_{1/2}$(GT)], as shown in Fig. \ref{fig7.1}. Note that the half-lives of $^{80g+m}$Ga do not decrease in magnitude because the $\beta$-decay constant is dominated by the energy window rather than selection rules. The upper-limit half-life ratios of $\it{T}_{1/2}$(GT) and $\it{T}_{1/2}$(GT+FF) are 1.92(11) and 1.92(12) for $^{80g}$Ga and $^{80m}$Ga, respectively. Compared to our observation, the theoretical model, which combines the quasiparticle random-phase approximation (QRPA) for the GT part with an empirical spreading of the QRPA strength and the gross theory for the FF part, produces a larger reduction of the half-life of $^{80}$Ga when including FF transitions: the calculated $\it{T}_{1/2}$(GT) to $\it{T}_{1/2}$(GT+FF) ratio of $^{80}$Ga is 2.0$-$2.5; see Fig. 10 of Ref. \cite{PhysRevC.67.055802}.

\section{V. Conclusion}

In conclusion, the detailed $\beta$-delayed $\gamma$ spectroscopy of $^{80}$Ge has been presented, with 13 and 14 states $\beta$ populated by $^{80g}$Ga and $^{80m}$Ga, respectively, being reported for the first time. The upper-limit fractions of FF transitions contributing to the decays of $^{80g}$Ga and $^{80m}$Ga are (48.0 $\pm$ 2.7)$\%$ and (47.9 $\pm$ 3.3)$\%$ of the observed total $\beta$ transition intensities of (78.2 $\pm$ 2.5)$\%$ and (82.2 $\pm$ 3.7)$\%$, respectively, which are comparable to GT transitions. The half-lives of $^{80g}$Ga and $^{80m}$Ga decrease in accordance with the upper limits of (48.0 $\pm$ 6.0)$\%$ and (47.9$\pm$7.2)$\%$, respectively, when including FF transitions, in contrast to those by GT transitions only. For determination of the spin-parity of observed states, angular distributions or correlations of deexcitation $\gamma$ rays need to be measured in the future.

\section*{ACKNOWLEDGMENTS}

The authors thankfully acknowledge the work of the ALTO technical staff for the excellent operation of the ISOL source. R.L. acknowledges support by the China Scholarship Council under Grant No.201804910509 and by a KU Leuven postdoctoral scholarship. R.L. and S.E. thank Professor N. Yoshinaga for the discussion on the shell-model calculation. C.D., A. K. and L. A. A. received funding from European Union's Horizon 2020 research and innovation program under Grant Agreement No. 771036 (ERC CoG MAIDEN). Use of the PARIS modular array from the PARIS Collaboration and Ge detectors from the French-UK IN2P3-STFC Gamma Loan Pool are acknowledged.

\section*{DATA AVAILABILITY}
The data that support the findings of this article are not publicly available upon publication because it is not technically feasible and/or the cost of preparing, depositing, and hosting the data would be prohibitive within the terms of this research project. The data are available from the authors upon reasonable request.

\bibliography{References_spectroscopy}
\onecolumngrid
\end{document}